\documentclass[12pt,english,floatfix,superscriptaddress,aps,prd,preprint,showkeys,nofootinbib]{revtex4}
\usepackage{amsmath}
\usepackage{amssymb}
\usepackage{amsbsy}
\usepackage{amsfonts}
\usepackage{amsopn}
\usepackage{amstext}
\usepackage{graphicx}
\usepackage[english]{babel}
\usepackage{color}
\usepackage{slashed}
\usepackage{esint}
\usepackage[dvips]{epsfig}
\usepackage[dvips]{graphicx}
\usepackage{float}
\usepackage{units}
\usepackage{textcomp}
\usepackage{wasysym}
\usepackage{hyperref}
\usepackage{slashed}
\usepackage{orcidlink} 

\begin{document}

\title{Nonlocal Rarita--Schwinger theory}

\author{Fernando M. Belchior}
\email{belchior@fisica.ufc.br}
\affiliation{Universidade Federal do Ceará (UFC), Departamento de Física,
 Campus do Pici, Fortaleza- CE, C.P. 6030, 60455-760- Brazil}

\author{Roberto V. Maluf {\orcidlink{0000-0002-9952-4589}}}
\email{r.v.maluf@fisica.ufc.br}
\affiliation{Universidade Federal do Ceará (UFC), Departamento de Física,
 Campus do Pici, Fortaleza- CE, C.P. 6030, 60455-760- Brazil}

\begin{abstract}
In this paper, one constructs a nonlocal extension of the Rarita-Schwinger theory for spin-$3/2$ fermions. Two classes of analytic form factors are considered: scalar form factors $f(\Box)$ and Dirac-operator form factors $f(\slashed{\partial})$. The massless theory is treated together with a covariant nonlocal gauge fixing, which allows the propagator to be written directly in terms of the spin-$3/2$ projector. In the massive theory, we show that the free Rarita-Schwinger constraints remain intact for analytic form factors, so that the unphysical spin-$1/2$ sector does not become dynamical. For $f(\Box)$ the tensor-spinor structure of the propagator is the same as in the local theory, while the pole equation is deformed by the scalar form factor. For $f(\slashed{\partial})$ the physical modes obey a nonlocal Dirac-type equation, leading to modified dispersion relations that can be written explicitly for exponential form factors. We discuss the conditions under which the construction is ghost-free at the free level and identify the natural limitations that must be addressed before interactions are introduced.
\end{abstract}

\maketitle

\section{Introduction}

The Rarita-Schwinger (RS) field has long provided the standard covariant description of spin-$3/2$ degrees of freedom in relativistic quantum field theory \cite{RaritaSchwinger1941}. A special interest in this theme is due to the fact that spin-$3/2$ fields arise naturally as the gravitino in supergravity, where local supersymmetry enforces a consistent coupling to gravity\cite{FreedmanFerraraVN1976,DeserZumino1976,GrisaruPendletonVN1977,vanNieuw1981SupergravityReview}. In its free form, the RS Lagrangian describes a vector-spinor $\psi_\mu$ subject to constraints that eliminate unphysical spin-$1/2$ components, a structure clarified in early analyses of higher-spin wave equations \cite{MoldauerCase1956}. However, once interactions are introduced, the constraint algebra and causal propagation become highly nontrivial. It has been shown that minimally coupled charged spin-$3/2$ fields exhibit inconsistencies and superluminal propagation in external electromagnetic backgrounds \cite{JohnsonSudarshan1961,VeloZwanziger1861969,VeloZwanziger1881969}, with further demonstrations of quantization and causality problems in related formulations \cite{Hagen1971,Singh1973}. These issues have motivated systematic efforts to define consistent interaction vertices and supplementary conditions, including constraints on spin-$3/2$ couplings and the role of off-shell ambiguities \cite{NathEtemadiKimel1971,SinghHagen1974Fermion}.

Moreover, gauge-invariant formulations for massless higher-spin fermions have been developed, rendering the gauge symmetry of the RS field more geometric and transparent \cite{FangFronsdal1978}. At the same time, studies of more ambitious setups, such as hypergravity, have shown that interacting higher-spin theories encounter serious consistency barriers, helping to map out what is or is not possible \cite{AragoneDeser1979Hypergravity}. For the massive spin-$3/2$ case, as encountered in hadronic physics (e.g., the $\Delta$ resonance), a central issue is the construction of interaction terms that prevent unphysical spin-$1/2$ components from contributing to observable quantities. This has led to the development of modern consistent-coupling schemes and carefully constructed effective Lagrangian approaches \cite{BenmerroucheDavidsonMukhopadhyay1989,Pascalutsa1998,PascalutsaTimmermans1999,Pascalutsa2001Equivalence}. Subsequent work has refined the conditions required for causal (i.e., nonsuperluminal) propagation in electromagnetic and gravitational backgrounds, and has clarified which issues persist for generic massive, charged higher-spin fields \cite{DeserPascalutsaWaldron2000,DeserWaldron2002,PorratiRahman2009}. Complementary studies have addressed practical aspects, including renormalization, phenomenology, and the implementation of consistency conditions directly at the level of the equations of motion \cite{KaloshinLomov2007,KrebsEpelbaumMeissner2009}.

Recent work has revisited Rarita-Schwinger QED from a modern effective field theory perspective, showing that quantum effects and symmetry-breaking deformations can qualitatively modify the gauge sector. In particular, the possibility of radiatively induced photon mass generation has been analyzed within this framework \cite{Ghasemkhani:2024akc}. In parallel, Lorentz-violating extensions of the RS theory have been developed and investigated at the level of their dynamics and consistency \cite{Gomes:2022btc}. Closely related studies have further demonstrated that Carroll–Field–Jackiw–type terms can emerge in RS settings, both in the massless Abelian case \cite{Gomes:2024qya} and in non-Abelian generalizations \cite{Gomes:2023qkj}. These works highlight how spin-$\tfrac{3}{2}$ matter can generate parity-odd gauge structures through loop effects and effective couplings.

The purpose of the present work is more modest and more precise: we construct the free nonlocal Rarita-Schwinger theory and determine how the spin-$3/2$ projector, the subsidiary constraints and the pole structure are modified by the form factor. This is the necessary first step before one can address the more delicate problem of interactions. Our construction follows the form-factor prescription recently explored for spin-$1/2$ fields in Ref.~\cite{Nascimento:2025ngc}, but it also keeps track of the tensor-spinor structure that is special to the Rarita-Schwinger field.

This paper is organized as follows. In Sec.~\ref{s2}, we review the local Rarita-Schwinger model. In Sec.~\ref{s3}, we present its nonlocal extension, first in the massless case and then in the massive case. Finally, in Sec.~\ref{s4}, we summarize the results and discuss possible continuations. Throughout the work we use flat spacetime with metric $\eta_{\mu\nu}=\mathrm{diag}(+,-,-,-)$ and Clifford algebra $\{\gamma^\mu,\gamma^\nu\}=2\eta^{\mu\nu}$.

\section{Rarita-Schwinger field theory}\label{s2}

In this section, we review the Rarita-Schwinger (RS) field theory describing a relativistic spin-$3/2$ fermion. In $d=4$, the RS field is a vector-spinor $\psi_\mu(x)$, carrying both a Lorentz vector index $\mu$ and a Dirac (or Majorana) spinor index (here suppressed). The corresponding Lorentz representation is given by \cite{MoldauerCase1956,vanNieuw1981SupergravityReview}
\begin{align}
(1/2,1/2)\otimes\Big[(1/2,0)\oplus(0,1/2)\Big]
\;\cong\;
\Big[(1,1/2)\oplus(1/2,1)\Big]\;\oplus\;\Big[(1/2,0)\oplus(0,1/2)\Big],
\end{align}
This representation shows that a generic field $\psi_\mu$ contains both spin-$3/2$ and spurious spin-$1/2$ components. The field carries a Dirac spinor index, corresponding to four complex components in the standard representation. Since $\psi_\mu$ also transforms as a Lorentz vector, it contains $4\times4=16$ complex components before constraints are imposed. A massive spin-$3/2$ particle, however, carries only $2s+1=4$ physical polarizations. The role of the Rarita-Schwinger equations is precisely to eliminate the redundant vector-spinor components through the gamma-trace and transversality constraints.

In order to construct a Lagrangian for the Rarita-Schwinger field, it is convenient to introduce antisymmetrized products of gamma matrices, defined as
\begin{align}
\gamma^{\mu\nu} \equiv \gamma^{[\mu}\gamma^{\nu]}=\frac{1}{2}(\gamma^\mu\gamma^\nu-\gamma^\nu\gamma^\mu),
\label{gamma-two}
\end{align}
and
\begin{align}
\gamma^{\mu\nu\rho}
\equiv \gamma^{[\mu}\gamma^\nu\gamma^{\rho]}
=\frac{1}{2}\left(\gamma^\mu\gamma^\nu\gamma^\rho-\gamma^\rho\gamma^\nu\gamma^\mu\right)
=\gamma^\mu\gamma^\nu\gamma^\rho-\gamma^\mu\eta^{\nu\rho}+\gamma^\nu\eta^{\mu\rho}-\gamma^\rho\eta^{\mu\nu}.
\label{gamma-three}
\end{align}
Here the brackets denote unit-weight antisymmetrization, and we use the slash notation $\slashed{\partial}=\gamma^\mu\partial_\mu$. The elementary Clifford-algebra derivation of Eqs.~\eqref{gamma-two} and \eqref{gamma-three} is given in Appendix~\ref{gamma-identities}. With these definitions, the standard free massive RS Lagrangian in $d=4$ reads \cite{RaritaSchwinger1941}
\begin{align}
\mathcal{L}_{\rm RS}=\bar\psi_\mu\left(i\,\gamma^{\mu\rho\nu}\partial_\rho
- m\,\gamma^{\mu\nu}\right)\psi_\nu,
\label{MRS-Lagrangian}
\end{align}
where $\bar\psi_\mu \equiv \psi_\mu^\dagger \gamma^0$.
The variation of \eqref{MRS-Lagrangian} with respect to $\bar\psi_\mu$ leads us to the Euler-Lagrange equations
\begin{align}
\left(i\,\gamma^{\mu\rho\nu}\partial_\rho- m\,\gamma^{\mu\nu}
\right)\psi_\nu=0.
\label{MRS-EOM}
\end{align}

It is important to point out that a consistent spin-$3/2$ dynamics must ensure that only the spin-$3/2$ sector propagates. As discussed above, for a massive particle, the physical number of states is $2s+1=4$, while for a massless particle, only helicities $\pm 3/2$ propagate (2 states), and the theory acquires a gauge redundancy that removes the unphysical polarizations. If we contract \eqref{MRS-EOM} with $\gamma_\mu$ and use the constraints $\gamma\cdot\psi=0$ and $\partial\cdot\psi=0$ (see Appendix~A), we show that each component obeys a Dirac equation, namely
\begin{align}
(i\slashed{\partial}-m)\psi_\mu = 0,
\label{MRS-reduced}
\end{align}
which is precisely the covariant characterization of a massive spin-$3/2$ field. The covariant massive RS propagator can be written compactly in terms of a tensor-spinor structure that projects onto the physical spin-$3/2$ sector on shell (see Appendix~B). Explicitly, the massive RS propagator is given by
\begin{align}
S_{\mu\nu}(p)=-\frac{\slashed{p}+m}{p^2-m^2+i\varepsilon}
\left[\eta_{\mu\nu}
-\frac{1}{3}\gamma_\mu\gamma_\nu
-\frac{1}{3m}\left(\gamma_\mu p_\nu-\gamma_\nu p_\mu\right)
-\frac{2}{3m^2}p_\mu p_\nu\right].
\label{MRS-propagator}
\end{align}
When contracted with conserved, gamma-traceless external currents (the physically relevant case), the terms involving $p_\mu$ and $\gamma_\mu$ do not contribute, and the propagator effectively reduces to the spin-$3/2$ projector times the Dirac propagator. Moreover, the extra pieces that depend on $p_\mu$ and $\gamma_\mu$ ensure Lorentz covariance and correct constraint propagation.

An interesting limit involves the massless case $m\to 0$. The massless RS Lagrangian becomes
\begin{align}
\mathcal{L}_{m=0}=i\,\bar\psi_\mu\,\gamma^{\mu\rho\nu}\partial_\rho\psi_\nu.
\label{MLRS-Lagrangian}
\end{align}
We observe that this Lagrangian is invariant (up to a total derivative) under the fermionic gauge transformation $\delta\psi_\mu = \partial_\mu \epsilon$, where $\epsilon(x)$ is a spinor gauge parameter. In addition, this gauge symmetry is the mechanism that removes the unphysical spin-$1/2$ sector in the massless theory and ensures that only helicities $\pm 3/2$ propagate. From a practical viewpoint one usually fixes the gauge by imposing a covariant gauge condition such as $\gamma^\mu\psi_\mu = 0$ and/or $\partial^\mu\psi_\mu = 0$, analogous to Lorenz gauge in electrodynamics, leaving only the transverse, gamma-traceless components. In momentum space with $p^2=0$, the physical polarizations satisfy $p^\mu u_\mu(p) = 0$, $\gamma^\mu u_\mu(p)=0$, $\slashed{p}\,u_\mu(p)=0$, which projects onto helicity $\pm 3/2$ states. In momentum space, it is convenient to decompose the vector-spinor into irreducible spin components. A convenient covariant gauge fixing is the $\gamma$-trace gauge
\begin{equation}
\mathcal{L}_{\rm gf}=-\frac{1}{\xi}\,(\bar\psi\!\cdot\!\gamma)\, i\slashed\partial\,(\gamma\!\cdot\!\psi)
\label{MLRS-gf}
\end{equation}
with gauge parameter $\xi$. By adding this gauge fixing and using the spin projector defined in Appendix~B, the propagator in the transverse gamma-traceless sector can be written compactly as
\begin{equation}
S_{\mu\nu}(p)=\frac{i}{\slashed p+i\varepsilon}\,P^{3/2}_{\mu\nu}(p),
\label{MLRS-propagator}
\end{equation}
where $P^{3/2}_{\mu\nu}$ is the covariant spin-$3/2$ projector. This form makes the particle content transparent: the only physical pole is the massless pole $p^2=0$, while the remaining spin-$1/2$ components are gauge artifacts. In the gauge $\xi=1$, the propagator is transverse and gamma-traceless inside the physical subspace, namely $p^\mu S_{\mu\nu}(p)=0$ and $\gamma^\mu S_{\mu\nu}(p)=0$ when it acts on admissible external currents.

\section{Nonlocal Rarita-Schwinger field theory}\label{s3}

In this section, we will construct a nonlocal extension of the Rarita-Schwinger theory. Initially, let us discuss some basic ideas of nonlocality. A nonlocal field theory generalizes conventional local quantum field theory (QFT) by allowing interactions that are not confined to a single spacetime point. Instead, the fundamental objects are nonlocal operators acting on fields over finite or infinite distances in spacetime. This framework was first systematically explored by Hideki Yukawa in the early 1950s as a way to describe elementary particles with finite size and to remove ultraviolet (UV) divergences inherent in local point-particle theories \cite{Yukawa:1950eq}.

As a concrete example, let us consider a real scalar field $\phi(x)$. A typical nonlocal action for $\phi(x)$ takes the following form \cite{Efimov:1967pjn, Briscese:2015zfa}
\begin{align}
S = \int d^4 x \left[ -\frac{1}{2} \phi(x) \, \mathcal{F}(\Box) \, (\Box - m_{\phi}^2) \phi(x) - V(\phi) \right],
\end{align}
where $\Box = \partial^\mu \partial_\mu$ is the d'Alembertian operator, $m_{\phi}$ is the mass parameter, $V(\phi)$ is a local or nonlocal potential, and $\mathcal{F}(z)$ is an entire analytic function (e.g., an exponential form factor) chosen to ensure a ghost-free model. As a concrete example, we can choose the function
\begin{align}
\mathcal{F}(z) = \exp\left( \frac{z}{\Lambda} \right),
\end{align}
where $\Lambda$ represents the nonlocality (or UV cutoff) scale. The operator \(\mathcal{F}(\Box)\) can be understood via its Fourier representation:
\begin{align}
\mathcal{F}(\Box) \phi(x) = \int \frac{d^d k}{(2\pi)^d} \, \tilde{\phi}(k) \, \mathcal{F}(-k^2) \, e^{-ik\cdot x}.
\end{align}

In momentum space, the free propagator reads
\begin{align}
D(p^2) = \frac{i}{p^2 \mathcal{F}(-p^2) - m^2 \mathcal{F}(-p^2)} = \frac{i \exp(-p^2/\Lambda)}{p^2 - m^2},
\end{align}
(for the common exponential choice). The exponential suppression \(\exp(-p^2/\Lambda)\) for large \(|p|\) improves the ultraviolet behavior of loop integrals and may render certain models UV finite or super-renormalizable.

To construct a nonlocal spin-$\tfrac{1}{2}$ theory, one follows the approach of the form-factor prescription introduced in \cite{Nascimento:2025ngc}. The key idea is to deform the first-order kinetic operator by an entire function of the Dirac operator, without introducing additional poles. Therefore, a convenient nonlocal deformation for fermions is built from the Dirac operator, replacing
$\slashed{\partial}\mapsto \slashed{\partial}\,f(\slashed{\partial})$ with $f$ an entire form factor satisfying $f(0)=1$, and $f$ entire and typically chosen with no zeros. A prototype for free massive nonlocal spin-$\tfrac12$ theory is described by the Lagrangian
\begin{equation}
\mathcal{L}^{(1/2)}_{\rm NL}=
\bar\Psi\big(i\slashed{\partial}\,f(\slashed{\partial})-m\big)\Psi,
\label{NL-Dirac-L}
\end{equation}
which produces the following equation of motion
\begin{equation}
\big(i\slashed{\partial}\,f(\slashed{\partial})-m\big)\Psi=0.
\label{NL-Dirac-EOM}
\end{equation}
For the spinorial deformation, one introduces an operator-valued entire form factor
\begin{align}
f(\slashed{\partial}) = \sum_{n=0}^{\infty} b_n \left(\frac{\slashed{\partial}}{\Lambda}\right)^n,
\end{align}
where $b_0=1$ and $\Lambda$ is a nonlocality scale. In momentum space, one obtains the propagator
\begin{equation}
S_{1/2}(p)=\frac{i}{\slashed{p}\,f(\slashed{p})-m}=
\frac{i\big(\slashed{p}\,f(\slashed{p})+m\big)}{p^2 f^2(\slashed{p})-m^2}.
\label{NL-Dirac-prop}
\end{equation}
Since $f$ is entire and has no zeros, no additional particle poles are introduced beyond those dictated by the local part. Using $(\slashed{p})^2=p^2\mathbf{1}$, any analytic $f(\slashed{p})$ can be decomposed as
\begin{equation}
f(\slashed{p})=f_{\rm e}(p^2)\,\mathbf{1}+f_{\rm o}(p^2)\,\slashed{p},
\label{f-even-odd1}
\end{equation}
where $f_{\rm e}$ is a even function, while $f_{\rm o}$ is an odd function, both determined by $f$. Below, we employ this approach for both massless and massive nonlocal RS fields.

\subsection{Massless nonlocal RS model}

Once we have presented the concept of nonlocality in QFT, let us construct a nonlocal version of the RS model discussed in the previous section. We can reach this goal by employing two representative classes of form factors $f(\Box)$ and $f(\slashed{\partial})$. We first propose a massless nonlocal RS model described by the following Lagrangian
\begin{align}\label{NRS1}
    \mathcal{L}_{\rm NRS}= i\bar\psi_\mu \gamma^{\mu\rho\nu} \partial_\rho f(\Box)\psi_\nu
\end{align}
One notes that \eqref{NRS1} is invariant under the fermionic gauge symmetry $\delta \psi_\mu = \partial_\mu \epsilon(x)$, with spinor parameter $\epsilon(x)$. Like the local case, this symmetry removes the unphysical spin-$\tfrac12$ components, leaving only the two physical helicity states $\pm\tfrac32$. Due to the gauge symmetry, it is necessary to add a convenient covariant gauge fixing given by
\begin{equation}
\mathcal{L}_{\rm gf}=-\frac{1}{\xi}\,(\bar\psi\!\cdot\!\gamma)\, i\slashed\partial f(\Box)\,(\gamma\!\cdot\!\psi)
\label{MLNRS_gf_gamma_trace1}
\end{equation}
with gauge parameter $\xi$. Using the spin projector of Appendix~B, we obtain for $\xi=1$
\begin{equation}
S_{\mu\nu}(p)=\frac{i}{\slashed p\, f(p^2)+i\varepsilon}\,P^{3/2}_{\mu\nu}(p).
\label{MLNRS-propagator}
\end{equation}
Compared with the local RS propagator, the only modification in the physical sector is the multiplicative form factor $1/f(p^2)$. If $f(z)$ is entire and nonzero in the complex plane, the propagator contains no new pole beyond the physical massless pole. Thus, at the free level, the deformation is gauge invariant and does not introduce additional ghost or tachyonic degrees of freedom. The infrared condition $f(0)=1$ guarantees the smooth recovery of the local RS theory.

On the other hand, by following the approach of \cite{Nascimento:2025ngc}, we can introduce an operator-valued entire form factor given by
\begin{align}
f(\slashed{\partial}) = \sum_{n=0}^{\infty} b_n \left(\frac{\slashed{\partial}}{\Lambda}\right)^n,
\end{align}
where $b_0=1$, $\Lambda$ is a nonlocality scale and $f(z)$ is chosen to be entire and nonvanishing in the complex plane. With this form factor, the Lagrangian takes the form
\begin{align}\label{NRS2}
    \mathcal{L}_{\rm NRS}= i\bar\psi_\mu \gamma^{\mu\rho\nu} \partial_\rho f(\slashed{\partial})\psi_\nu
\end{align}
This Lagrangian is also invariant under the fermionic gauge symmetry. It can be conveniently written as
\begin{align}\label{NRS-L}
\mathcal{L}_{\rm NRS}= \,\bar\psi_\mu \Lambda^{\mu\nu} \,\psi_\nu,
\end{align}
where $\Lambda^{\mu\nu}$ is the wave operator given by
\begin{align}
\Lambda^{\mu\nu}=i\slashed{\partial}f(\slashed{\partial})\eta^{\mu\nu}-i(\gamma^\mu \partial^\nu+\gamma^\nu \partial^\mu)f(\slashed{\partial})
+i\gamma^\mu\slashed{\partial}f(\slashed{\partial})\gamma^\nu.
\end{align}
We now add the convenient covariant $\gamma$-trace gauge-fixing term
\begin{equation}
\mathcal{L}_{\rm gf}=-\frac{1}{\xi}\,(\bar\psi\!\cdot\!\gamma)\, i\slashed\partial f(\slashed{\partial})\,(\gamma\!\cdot\!\psi)
\label{MLNRS_gf_gamma_trace}
\end{equation}
with gauge parameter $\xi$. Using the same spin-projector decomposition, we obtain for $\xi=1$ the propagator
\begin{equation}
S_{\mu\nu}(p)=\frac{i}{\slashed p\,f(\slashed p)+i\varepsilon}\,P^{3/2}_{\mu\nu}(p).
\label{MLNRS-propagator-slash}
\end{equation}
This result should be read as an operator equation in spinor space. Since $f(\slashed p)$ is an analytic function of $\slashed p$, it commutes with $\slashed p$, and the inversion proceeds as in the local theory after the replacement $\slashed p\rightarrow \slashed p f(\slashed p)$. The difference from the $f(\Box)$ deformation is that the form factor is now a genuine matrix in spinor space. Consequently, its even and odd parts may affect not only the overall ultraviolet damping but also the detailed pole equation of the spinorial denominator.

\subsection{Massive nonlocal RS model}

For the massive case, a direct massive nonlocal spin-$\tfrac32$ theory has the Lagrangian
\begin{align}
\mathcal{L}=\bar\psi_\mu\left(i\,\gamma^{\mu\rho\nu}\partial_\rho f(\Box)- m\,\gamma^{\mu\nu}\right)\psi_\nu.
\label{MNRS-Lagrangian1}
\end{align}
The variation of this Lagrangian with respect to $\bar\psi_\mu$ leads us to the following Euler-Lagrange equation
\begin{align}
\left(i\,\gamma^{\mu\rho\nu}\partial_\rho f(\Box)- m\,\gamma^{\mu\nu}
\right)\psi_\nu=0.
\label{MNRS-EOM1}
\end{align}
It is important to observe that for $m\neq 0$, the nonlocal equation \eqref{MNRS-EOM1} still implies $\gamma\cdot\psi=0$ and $p\cdot\psi=0$, so no extra propagating spin-$\tfrac12$ component is forced to appear at the free level. With \eqref{RS-NL-constraints}, the physical spin-$\tfrac32$ sector obeys a nonlocal Dirac-type equation:
\begin{equation}
\big(\slashed{p}\,f(p^2)-m\big)\psi_\mu(p)=0.
\label{MNRS-reduced1}
\end{equation}

In momentum space, the propagator for this model reads
\begin{equation}
S^{(3/2)}_{\mu\nu}(p)=-\frac{\slashed{p}f(p^2)+m}{p^2f^2(p^2)-m^2+i\varepsilon}
\left[\eta_{\mu\nu}-\frac{1}{3}\gamma_\mu\gamma_\nu-\frac{1}{3m}\left(\gamma_\mu p_\nu-\gamma_\nu p_\mu\right)-\frac{2}{3m^2}p_\mu p_\nu\right].
\label{MNRS-prop1}
\end{equation}

Additionally, a spin-$\tfrac32$ analogue of the Dirac-operator deformation is obtained by replacing $\partial_\nu\mapsto \partial_\nu f(\slashed{\partial})$ in the kinetic term, so that the Lagrangian \eqref{MRS-Lagrangian} takes the form 
\begin{align}
\mathcal{L}=\bar\psi_\mu\left(i\,\gamma^{\mu\rho\nu}\partial_\rho f(\slashed{\partial})- m\,\gamma^{\mu\nu}
\right)\psi_\nu,
\label{MNRS-Lagrangian2}
\end{align}
The variation of this Lagrangian with respect to $\bar\psi_\mu$ leads us to the Euler-Lagrange equations
\begin{align}
\left(i\,\gamma^{\mu\rho\nu}\partial_\rho f(\slashed{\partial})-m\,\gamma^{\mu\nu}\right)\psi_\nu=0.
\label{RS-EOM2}
\end{align}

We rewrite the Lagrangian as
\begin{align}
\mathcal{L}_{\rm NRS}= \,\bar\psi_\mu \Lambda^{\mu\nu} \,\psi_\nu,
\label{RS-NL-L}
\end{align}
where 
\begin{align}
\Lambda^{\mu\nu}=(i\slashed{\partial}f(\slashed{\partial})-m)\eta^{\mu\nu}-i(\gamma^\mu \partial^\nu+\gamma^\nu \partial^\mu)f(\slashed{\partial})
+i\gamma^\mu(\slashed{\partial}f(\slashed{\partial})+m)\gamma^\nu.
\end{align}

Like the previous case, we show in Appendix~A that for $m\neq 0$ the nonlocal equations still imply $\gamma\cdot\psi=0$ and $p\cdot\psi=0$, so no extra propagating spin-$\tfrac12$ component is forced to appear at the free level. With \eqref{RS-NL-constraints}, the physical spin-$\tfrac32$ sector obeys a nonlocal Dirac-type equation:
\begin{equation}
\big(\slashed{p}\,f(\slashed{p})-m\big)\psi_\mu(p)=0.
\label{MNRS-reduced2}
\end{equation}
Therefore the dispersion relation for the propagating spin-$\tfrac32$ modes is inherited from the determinant condition for $\slashed{p}\,f(\slashed{p})-m$ (with multiplicity corresponding to the four massive spin states). If the function $f$ is entire and has no zeros, the pole content is not enlarged relative to the local RS theory. 

A common class of nonlocal functions is exponential form factors built from $\slashed{\partial}/\Lambda$. For instance, we can use the function
\begin{equation}
f_I(\slashed{\partial})=e^{-\slashed{\partial}/\Lambda},
\label{fI}
\end{equation}
which reduces to the local theory as $\Lambda\to\infty$. In momentum space, this function yields a non-polynomial deformation of the on-shell condition determined by
\begin{equation}
\det\big(\slashed{p}\,f_I(\slashed{p})-m\big)=0,
\end{equation}
where the determinant is taken in spinor space. This equation can be expressed in closed form (for suitable branches) using the Lambert $W$ function. Another suitable choice is
\begin{equation}
f_{II}(\slashed{\partial})=e^{-i\slashed{\partial}/\Lambda},
\label{fII}
\end{equation}
which leads to oscillatory operator structures and correspondingly modified dispersion relations. In the massive nonlocal Rarita-Schwinger (RS) model built with a Dirac-operator form factor, the free constraints imply that the physical spin-$\tfrac32$ components obey a nonlocal Dirac-type equation, so the dispersion relation of propagating RS modes is determined by the same spinorial matrix equation as in the spin-$\tfrac12$ case, up to multiplicities. Because $f(\slashed{p})$ is a (matrix) function of $\slashed{p}$, it commutes with $\slashed{p}$. Then, multiplying \eqref{MNRS-reduced2} on the left by $\slashed{p}\,f(\slashed{p})+m$, we get
\begin{equation}
\big(\slashed{p}\,f(\slashed{p})+m\big)\big(\slashed{p}\,f(\slashed{p})-m\big)\psi_\mu=
\big(\slashed{p}^{\,2}\,f^2(\slashed{p})-m^2\big)\psi_\mu=0.
\label{RS_NL_second_order_raw}
\end{equation}
Proceeding further, we can use $\slashed{p}^{\,2}=p_\mu p^\mu\,\hat I \equiv p^2\hat I$, thereby writing
\begin{equation}
\big(p^2 f^2(\slashed{p})-m^2\big)\psi_\mu=0.
\label{RS_NL_second_order}
\end{equation}

To make the dispersion relation explicit, it is useful to decompose the matrix function $f^2(\slashed p)$ in the basis $\{\hat I,\slashed p\}$. We write
\begin{equation}
f^2(\slashed p)\equiv g(p)=a_1(p)\,\hat I+a_2(p)\,\slashed p,
\label{g_decomposition}
\end{equation}
where $a_1$ and $a_2$ are scalar functions of the Lorentz invariant $p^2$. Equation~\eqref{RS_NL_second_order} then becomes
\begin{equation}
S(p)\psi_\mu(p)=0,\qquad
S(p)=\big(p^2a_1(p)-m^2\big)\hat I+p^2a_2(p)\slashed p .
\label{eq:S_matrix_def}
\end{equation}
Nontrivial solutions require $\det S(p)=0$. Since $(\slashed p)^2=p^2\hat I$, the determinant of a matrix of the form $A\hat I+B\slashed p$ is
\begin{equation}
\det\big(A\hat I+B\slashed p\big)=\big(A^2-B^2p^2\big)^2.
\end{equation}
Therefore the pole equation for the physical spin-$\tfrac32$ excitations is
\begin{equation}
\big[p^2a_1(p)-m^2\big]^2-p^6a_2^2(p)=0,
\end{equation}
or, equivalently,
\begin{equation}
p^2a_1(p)-m^2=\pm\,\big(p^2\big)^{3/2}a_2(p).
\label{dispersion_general}
\end{equation}
This is the non-polynomial dispersion relation for each physical spin-$\tfrac32$ mode. It is the spin-$3/2$ analogue of the dispersion equation obtained in the nonlocal spin-$1/2$ model, with the multiplicity enlarged by the vector-spinor polarizations.

As a first example, consider
\begin{equation}
f_I(\slashed\partial)=\exp(-\slashed\partial/\Lambda).
\end{equation}
In momentum space we introduce $\rho\equiv\sqrt{p^2}=\sqrt{E^2-|\vec p|^2}$ for timelike momenta. The square of the form factor can be written as
\begin{align}
a_1(p)&=\cosh\!\left(\frac{2\rho}{\Lambda}\right),
&
a_2(p)&=-\frac{1}{\rho}\sinh\!\left(\frac{2\rho}{\Lambda}\right).
\label{a1a2_fI}
\end{align}
Substitution into \eqref{dispersion_general} gives
\begin{equation}
\rho^2\exp\!\left(\pm\frac{2\rho}{\Lambda}\right)=m^2.
\label{dispersion_fI}
\end{equation}
For the branch with the plus sign, the solution can be expressed through the principal Lambert function,
\begin{equation}
\rho=\Lambda W_0\!\left(\frac{m}{\Lambda}\right),
\label{Lambert_solution}
\end{equation}
which yields, for $\Lambda\gg m$,
\begin{equation}
E^2=|\vec p|^2+m^2-\frac{2m^3}{\Lambda}+\mathcal{O}\!\left(\frac{1}{\Lambda^2}\right).
\label{IR_expansion}
\end{equation}
The opposite sign corresponds to $\rho=-\Lambda W(-m/\Lambda)$ and is real only in the domain where the appropriate Lambert-$W$ branch is real. This illustrates a general feature of Dirac-operator form factors: even if the function is entire, the physical branch structure must be chosen with care.

As a second example, take
\begin{equation}
f_{II}(\slashed\partial)=\exp(-i\slashed\partial/\Lambda).
\end{equation}
In this case,
\begin{align}
a_1(p)&=\cos\!\left(\frac{2\rho}{\Lambda}\right),
&
a_2(p)&=-\frac{i}{\rho}\sin\!\left(\frac{2\rho}{\Lambda}\right),
\label{a1a2_fII}
\end{align}
and the dispersion equation becomes
\begin{equation}
\rho^2\cos\!\left(\frac{2\rho}{\Lambda}\right)-m^2
=\mp i\rho^2\sin\!\left(\frac{2\rho}{\Lambda}\right).
\label{dispersion_fII_raw}
\end{equation}
This relation is generically complex, reflecting the oscillatory character of the form factor. A real and unitary spectrum therefore requires additional restrictions on the allowed branches, the reality properties of the action, or the choice of form factor. For this reason, exponentially damped entire functions are usually the safer option when the goal is a ghost-free effective theory.

\section{Final remarks}\label{s4}

In this work, we have constructed a nonlocal free Rarita-Schwinger theory by introducing analytic form factors in the kinetic operator and by following the consequences of this deformation through the constraints, propagators, and dispersion relations. The main point of the analysis is that nonlocality can be incorporated without spoiling the spin-$3/2$ structure of the free theory. In both the local and nonlocal models, the gamma-trace and transversality constraints remove the unphysical spin-$1/2$ components, so the vector-spinor field propagates the expected Rarita-Schwinger degrees of freedom.

For scalar form factors $f(\Box)$, the result is particularly transparent. In the massless model the propagator is the usual spin-$3/2$ projector multiplied by the nonlocal factor $1/f(p^2)$. If $f$ is entire and has no zeros, no new massless poles are generated. In the massive model, the tensor-spinor structure of the propagator remains the same as in the local theory, while the physical pole is governed by the deformed equation $p^2f^2(p^2)-m^2=0$. Thus, the deformation acts as a controlled modification of the dispersion relation rather than as a source of additional tensor-spinor states. For Dirac-operator form factors $f(\slashed\partial)$, the structure is richer. Since the form factor is a matrix in spinor space, its even and odd parts contribute differently to the spinorial denominator. Nevertheless, after the Rarita-Schwinger constraints are imposed, the physical spin-$3/2$ modes obey a nonlocal Dirac-type equation. This allowed us to derive a compact determinant condition and to obtain explicit dispersion relations for exponential choices of $f$. The exponentially damped form factor gives a real branch expressible in terms of the Lambert-$W$ function and smoothly reproduces the local dispersion relation when $\Lambda\rightarrow\infty$. The oscillatory example, on the other hand, shows that analyticity alone is not enough: reality, unitarity and branch selection are additional physical requirements.

The present construction should be regarded as the free-field foundation for a broader program. The next natural step is the inclusion of interactions. This is delicate because interacting massive spin-$3/2$ fields already face constraint and causality problems in local field theory. In a nonlocal setting, one must also control the ordering of form factors, the definition of gauge-covariant nonlocal operators, the analytic structure of retarded Green functions and the compatibility between nonlocality and the Rarita-Schwinger subsidiary constraints. A consistent interacting model should therefore be built as an effective field theory below the nonlocality scale $\Lambda$, with couplings restricted by the requirement that no unphysical spin-$1/2$ mode becomes dynamical.

In this sense, the results obtained here provide a clean starting point. They show that a ghost-free, ultraviolet-soft nonlocal deformation of the Rarita-Schwinger field is possible at the free level, and they identify the precise points where further consistency conditions enter. Future work may develop the gauge-covariant version of the theory, study quantum corrections induced by spin-$3/2$ fields, analyze causal propagation in external backgrounds, and explore whether nonlocality can soften the ultraviolet behavior of Rarita-Schwinger effective theories without spoiling their constrained higher-spin structure.

\appendix

\section{Derivation of the gamma-matrix identities in Eqs. (2) and (3)}
\label{gamma-identities}

In this appendix, we spell out the short algebraic derivation of the gamma-matrix relations used in the construction of the Rarita-Schwinger operator. The only input is the Clifford algebra
\begin{equation}
\{\gamma^\mu,\gamma^\nu\}=\gamma^\mu\gamma^\nu+\gamma^\nu\gamma^\mu=2\eta^{\mu\nu}\mathbf{1}.
\label{Clifford-app}
\end{equation}
We use unit-weight antisymmetrization, so that for two objects one has
\begin{equation}
A^{[\mu}B^{\nu]}\equiv \frac{1}{2}\left(A^\mu B^\nu-A^\nu B^\mu\right).
\label{antisym-two-app}
\end{equation}
Taking $A^\mu=B^\mu=\gamma^\mu$, this immediately gives
\begin{equation}
\gamma^{[\mu}\gamma^{\nu]}
=\frac{1}{2}\left(\gamma^\mu\gamma^\nu-\gamma^\nu\gamma^\mu\right)
\equiv \gamma^{\mu\nu},
\end{equation}
which is Eq.~\eqref{gamma-two}. Notice also that the Clifford algebra allows the useful decomposition
\begin{equation}
\gamma^\mu\gamma^\nu=\eta^{\mu\nu}\mathbf{1}+\gamma^{\mu\nu}.
\label{gamma-product-decomp-app}
\end{equation}
Thus the symmetric part of the product is fixed by the metric, while the antisymmetric part is precisely $\gamma^{\mu\nu}$. For three gamma matrices the unit-weight antisymmetrized product is
\begin{align}
\gamma^{[\mu}\gamma^\nu\gamma^{\rho]}=\frac{1}{6}\Big(&\gamma^\mu\gamma^\nu\gamma^\rho
+\gamma^\nu\gamma^\rho\gamma^\mu
+\gamma^\rho\gamma^\mu\gamma^\nu 
\nonumber\\
&-\gamma^\mu\gamma^\rho\gamma^\nu
-\gamma^\nu\gamma^\mu\gamma^\rho
-\gamma^\rho\gamma^\nu\gamma^\mu\Big).
\label{antisym-three-start-app}
\end{align}
The expression can be reduced by moving all factors to the ordered product $\gamma^\mu\gamma^\nu\gamma^\rho$ and using $\gamma^\alpha\gamma^\beta=2\eta^{\alpha\beta}-\gamma^\beta\gamma^\alpha$. The needed reorderings are
\begin{align}
\gamma^\nu\gamma^\rho\gamma^\mu
&=\gamma^\mu\gamma^\nu\gamma^\rho
+2\eta^{\mu\rho}\gamma^\nu
-2\eta^{\mu\nu}\gamma^\rho,
\label{eq:reorder1-app}\\
\gamma^\rho\gamma^\mu\gamma^\nu
&=\gamma^\mu\gamma^\nu\gamma^\rho
+2\eta^{\mu\rho}\gamma^\nu
-2\eta^{\nu\rho}\gamma^\mu,
\label{eq:reorder2-app}\\
\gamma^\mu\gamma^\rho\gamma^\nu
&=2\eta^{\nu\rho}\gamma^\mu-
\gamma^\mu\gamma^\nu\gamma^\rho,
\label{reorder3-app}\\
\gamma^\nu\gamma^\mu\gamma^\rho
&=2\eta^{\mu\nu}\gamma^\rho-
\gamma^\mu\gamma^\nu\gamma^\rho,
\label{reorder4-app}\\
\gamma^\rho\gamma^\nu\gamma^\mu
&=2\eta^{\mu\nu}\gamma^\rho
-2\eta^{\mu\rho}\gamma^\nu
+2\eta^{\nu\rho}\gamma^\mu
-\gamma^\mu\gamma^\nu\gamma^\rho.
\label{reorder5-app}
\end{align}
Substituting these relations into Eq.~\eqref{antisym-three-start-app}, all terms combine into
\begin{equation}
\gamma^{[\mu}\gamma^\nu\gamma^{\rho]}
=\gamma^\mu\gamma^\nu\gamma^\rho
-\eta^{\nu\rho}\gamma^\mu
+\eta^{\mu\rho}\gamma^\nu
-\eta^{\mu\nu}\gamma^\rho.
\label{gamma-three-derived-app}
\end{equation}
Since the metric is a scalar in spinor space, Eq.~\eqref{gamma-three-derived-app} is exactly the last equality in Eq.~\eqref{gamma-three}. It is also useful to verify the compact form displayed in the same equation. From Eq.~\eqref{reorder5-app}, one obtains
\begin{align}
\frac{1}{2}\left(\gamma^\mu\gamma^\nu\gamma^\rho-\gamma^\rho\gamma^\nu\gamma^\mu\right)
&=\gamma^\mu\gamma^\nu\gamma^\rho
-\eta^{\mu\nu}\gamma^\rho
+\eta^{\mu\rho}\gamma^\nu
-\eta^{\nu\rho}\gamma^\mu,
\end{align}
which is the same result after a trivial rearrangement of the metric factors. This completes the derivation of Eq.~\eqref{gamma-three}.

\section{Constraints for local and nonlocal RS models}

In this appendix, we derive the constraints for the local and nonlocal RS models. We first act with $\partial_\mu$ on \eqref{MRS-EOM} as follows
\begin{equation}
i\,\partial_\mu\gamma^{\mu\nu\rho}\partial_\nu\psi_\rho
- m\,\partial_\mu\gamma^{\mu\rho}\psi_\rho=0.
\end{equation}
Observe that the first term vanishes because $\gamma^{\mu\nu\rho}$ is antisymmetric in $\mu\nu$ while $\partial_\mu\partial_\nu$ is symmetric, then one has
\begin{equation}
\partial_\mu\gamma^{\mu\nu\rho}\partial_\nu\psi_\rho=
\gamma^{\mu\nu\rho}\partial_\mu\partial_\nu\psi_\rho=0.
\end{equation}
Hence, we have
\begin{equation}
\gamma^{\mu\rho}\partial_\mu\psi_\rho=0.
\label{RS-div-relation-raw}
\end{equation}
Using $\gamma^{\mu\rho}=\gamma^\mu\gamma^\rho-\eta^{\mu\rho}$, we rewrite
\begin{equation}
\gamma^{\mu\rho}\partial_\mu\psi_\rho=
\gamma^\mu\partial_\mu(\gamma^\rho\psi_\rho)-\partial^\rho\psi_\rho
=\slashed{\partial}(\gamma\cdot\psi) - \partial\cdot\psi,
\end{equation}
so \eqref{RS-div-relation-raw} becomes
\begin{equation}
\partial\cdot\psi=\slashed{\partial}(\gamma\cdot\psi).
\label{RS-div-rel}
\end{equation}

We next contract \eqref{MRS-EOM} with $\gamma_\mu$. In $d$ dimensions, one has the following identities
\begin{equation}
\gamma_\mu \gamma^{\mu\nu\rho} = (d-2)\,\gamma^{\nu\rho},
\qquad
\gamma_\mu \gamma^{\mu\rho} = (d-1)\,\gamma^\rho,
\end{equation}
so in $d=4$, one obtains
\begin{equation}
\gamma_\mu \gamma^{\mu\nu\rho} = 2\,\gamma^{\nu\rho},
\qquad
\gamma_\mu \gamma^{\mu\rho} = 3\,\gamma^\rho.
\end{equation}
Therefore, the gamma-trace of \eqref{MRS-EOM} gives
\begin{equation}
2i\,\gamma^{\nu\rho}\partial_\nu\psi_\rho - 3m\,\gamma^\rho\psi_\rho = 0.
\label{RS-gamma-trace-raw}
\end{equation}
But
\begin{equation}
\gamma^{\nu\rho}\partial_\nu\psi_\rho=(\gamma^\nu\gamma^\rho-\eta^{\nu\rho})\partial_\nu\psi_\rho=\slashed{\partial}(\gamma\cdot\psi)-\partial\cdot\psi,
\end{equation}
so \eqref{RS-gamma-trace-raw} becomes
\begin{equation}
2i\Big[\slashed{\partial}(\gamma\cdot\psi)-\partial\cdot\psi\Big]
-3m\,(\gamma\cdot\psi)=0.
\label{RS-gamma-trace}
\end{equation}

Insert \eqref{RS-div-rel} into \eqref{RS-gamma-trace}, we obtain
\begin{equation}
2i\Big[\slashed{\partial}(\gamma\cdot\psi)-\slashed{\partial}(\gamma\cdot\psi)\Big]-3m\,(\gamma\cdot\psi)=0
\quad\Rightarrow\quad
\gamma\cdot\psi=0 \quad (m\neq 0).
\end{equation}
Then \eqref{RS-div-rel} implies $\partial\cdot\psi=0$.
Thus the RS equations dynamically enforce the pure spin-$3/2$ constraints.
We now verify whether the constraints obtained previously still hold in the nonlocal case, working directly in position space. Starting from the nonlocal equations of motion
\begin{equation}
\Big(i\,\gamma^{\mu\nu\rho}\partial_\nu\, f(\slashed{\partial})-m\,\gamma^{\mu\rho}\Big)\psi_\rho(x)=0,
\label{RS-NL-EOM}
\end{equation}
we contract with $i\partial_\mu$. This yields
\begin{equation}
i\gamma^{\mu\nu\rho}\partial_\mu\partial_\nu\, f(\slashed{\partial})\,\psi_\rho
-m\partial_\mu\gamma^{\mu\rho}\psi_\rho=0.
\end{equation}
Again, since $\partial_\mu\partial_\nu$ is symmetric while $\gamma^{\mu\nu\rho}$ is antisymmetric in $(\mu,\nu)$, one has
\begin{equation}
\gamma^{\mu\nu\rho}\partial_\mu\partial_\nu=0,
\end{equation}
and therefore
\begin{equation}
\gamma^{\mu\rho}\partial_\mu\psi_\rho=0
\quad\Longleftrightarrow\quad
\slashed{\partial}\,(\gamma\!\cdot\!\psi)-(\partial\!\cdot\!\psi)=0.
\label{RS-NL-div}
\end{equation}

Next, we contract \eqref{RS-NL-EOM} with $\gamma_\mu$ gives
\begin{equation}
\gamma_\mu\gamma^{\mu\nu\rho}(i\partial_\nu)\, f(\slashed{\partial})\,\psi_\rho
-m\,\gamma_\mu\gamma^{\mu\rho}\psi_\rho=0.
\end{equation}
Using the $d=4$ identities
\begin{equation}
\gamma_\mu\gamma^{\mu\nu\rho}=2\gamma^{\nu\rho},
\qquad
\gamma_\mu\gamma^{\mu\rho}=3\gamma^\rho,
\qquad
\gamma^{\nu\rho}(i\partial_\nu)=i\slashed{\partial}\gamma^\rho-i\partial^\rho,
\end{equation}
we obtain
\begin{equation}
2\big(i\slashed{\partial}\gamma^\rho-i\partial^\rho\big)\,f(\slashed{\partial})\,\psi_\rho
-3m\,(\gamma\!\cdot\!\psi)=0.
\label{RS-NL-trace}
\end{equation}

To handle the non-commutativity between $\gamma^\rho$ and $f(\slashed{\partial})$, we need to use the standard even/odd
decomposition (valid for analytic $f$)
\begin{equation}
f(\slashed{\partial})=f_{\rm e}(\Box)\,\mathbf{1}+f_{\rm o}(\Box)\,\slashed{\partial},
\label{f-even-odd}
\end{equation}
together with the Clifford identity
\begin{equation}
\gamma^\rho\slashed{\partial}=2\partial^\rho-\slashed{\partial}\gamma^\rho.
\label{gamma-slash-identity}
\end{equation}
First, since $[\partial^\rho,f(\slashed{\partial})]=0$ for constant-coefficient pseudodifferential operators, we have
\begin{equation}
i\partial^\rho f(\slashed{\partial})\psi_\rho=
f(\slashed{\partial})\,i(\partial\!\cdot\!\psi).
\label{pfpsi}
\end{equation}
Next,
\begin{align}
\gamma^\rho f(\slashed{\partial})\psi_\rho
&=f_{\rm e}(\Box)(\gamma\!\cdot\!\psi)+f_{\rm o}(\Box)\gamma^\rho\slashed{\partial}\psi_\rho \nonumber\\
&=f_{\rm e}(\Box)(\gamma\!\cdot\!\psi)+f_{\rm o}(\Box)\big(2(\partial\!\cdot\!\psi)-\slashed{\partial}(\gamma\!\cdot\!\psi)\big).
\label{gammafpsi}
\end{align}
Using \eqref{RS-NL-div} to replace $\partial\!\cdot\!\psi=\slashed{\partial}(\gamma\!\cdot\!\psi)$, the bracket becomes
$2\slashed{\partial}(\gamma\!\cdot\!\psi)-\slashed{\partial}(\gamma\!\cdot\!\psi)=\slashed{\partial}(\gamma\!\cdot\!\psi)$, hence
\begin{equation}
\gamma^\rho f(\slashed{\partial})\psi_\rho
=\big(f_{\rm e}(\Box)+f_{\rm o}(\Box)\slashed{\partial}\big)(\gamma\!\cdot\!\psi)
=f(\slashed{\partial})(\gamma\!\cdot\!\psi).
\label{gammafpsi-simpl}
\end{equation}

Inserting \eqref{pfpsi} and \eqref{gammafpsi-simpl} into \eqref{RS-NL-trace}, one obtains
\begin{align}
0&=2\Big(i\slashed{\partial}\,f(\slashed{\partial})(\gamma\!\cdot\!\psi)
-f(\slashed{\partial})\,i(\partial\!\cdot\!\psi)\Big)
-3m(\gamma\!\cdot\!\psi)\nonumber\\
&=2 f(\slashed{\partial})\Big(i\slashed{\partial}(\gamma\!\cdot\!\psi)-i(\partial\!\cdot\!\psi)\Big)-3m(\gamma\!\cdot\!\psi).
\end{align}
The bracket vanishes by \eqref{RS-NL-div}. Therefore, for $m\neq0$, we have
\begin{equation}
\gamma\!\cdot\!\psi=0,
\qquad
\partial\!\cdot\!\psi=0.
\label{RS-NL-constraints}
\end{equation}

\section{Spin projectors and RS propagator}

In this appendix, we calculate the RS propagator by employing a suitable basis of spin projectors, which provides a clean way to separate spin-$3/2$ from spin-$1/2$ components in momentum space. We start by defining the transverse and longitudinal tensor projectors
\begin{equation}
\theta_{\mu\nu}\equiv \eta_{\mu\nu}-\frac{p_\mu p_\nu}{p^2},
\qquad
\omega_{\mu\nu}\equiv \frac{p_\mu p_\nu}{p^2},
\qquad
\theta^2=\theta,\ \omega^2=\omega,\ \theta\omega=0,
\label{thetaomega}
\end{equation}
These projectors obey the algebra $\theta^2=\theta,\ \omega^2=\omega,\ \theta\omega=0$. Besides, we define the gamma-transverse vector
\begin{equation}
\hat\gamma_\mu \equiv \theta_{\mu\nu}\gamma^\nu
=\gamma_\mu-\frac{p_\mu\slashed{p}}{p^2},
\label{hatgamma}
\end{equation}
which obeys $p^\mu\hat\gamma_\mu=0$. Then, a standard complete set of spin projectors is given by
\begin{align}
P^{3/2}_{\mu\nu}(p)
&\equiv
\theta_{\mu\nu}-\frac{1}{3}\hat\gamma_\mu\hat\gamma_\nu,
\label{P32}\\
P^{1/2}_{11\,\mu\nu}(p)
&\equiv
\frac{1}{3}\hat\gamma_\mu\hat\gamma_\nu,
\qquad
P^{1/2}_{22\,\mu\nu}(p)
\equiv
\omega_{\mu\nu},
\label{P12diag}\\
P^{1/2}_{12\,\mu\nu}(p)
&\equiv
\frac{1}{\sqrt{3}}\hat\gamma_\mu\,\frac{p_\nu}{\sqrt{p^2}},
\qquad
P^{1/2}_{21\,\mu\nu}(p)
\equiv
\frac{1}{\sqrt{3}}\frac{p_\mu}{\sqrt{p^2}}\,\hat\gamma_\nu.
\label{P12off}
\end{align}
They satisfy the algebra (matrix-unit structure in the spin-$1/2$ sector)
\begin{equation}
P^{3/2}P^{3/2}=P^{3/2},\qquad
P^{3/2}P^{1/2}_{ij}=0=P^{1/2}_{ij}P^{3/2},
\qquad
P^{1/2}_{ij}P^{1/2}_{kl}=\delta_{jk}P^{1/2}_{il},
\label{proj-algebra}
\end{equation}
and completeness
\begin{equation}
P^{3/2}_{\mu\nu}+P^{1/2}_{11\,\mu\nu}+P^{1/2}_{22\,\mu\nu}=\eta_{\mu\nu}.
\label{proj-complete}
\end{equation}

The RS operator from \eqref{MRS-Lagrangian} reads
\begin{equation}
\Lambda^{\mu\nu}(p)=\gamma^{\mu\rho\nu}p_\rho - m\,\gamma^{\mu\nu},
\end{equation}
or, equivalently
\begin{equation}
\Lambda^{\mu\nu}(p)=(\slashed{p}-m)\eta^{\mu\nu}-\gamma^\mu p^\nu-\gamma^\nu p^\mu+\gamma^\mu(\slashed{p}+m)\gamma^\nu.
\label{Lambda-expanded-again}
\end{equation}

The operator $\Lambda^{\mu\nu}$ is invertible as a $16\times 16$ matrix in vector-spinor space for $p^2\neq m^2$. First, we isolate the spin-$3/2$ subspace with $p\cdot\psi=0=\gamma\cdot\psi$, on which $\Lambda^{\mu\nu}$ reduces to $(\slashed{p}-m)\delta^\mu{}_\nu$. Then, we fix the remaining spin-$1/2$ sector by demanding full inversion $\Lambda^{\mu\rho}S_{\rho\nu}=i\delta^\mu{}_\nu$. After some calculations, we arrive at the standard covariant RS propagator given by
\begin{equation}
S_{\mu\nu}(p)=-\frac{i(\slashed{p}+m)}{p^2-m^2+i\varepsilon}
\left[\eta_{\mu\nu}-\frac{1}{3}\gamma_\mu\gamma_\nu
-\frac{1}{3m}\left(\gamma_\mu p_\nu-\gamma_\nu p_\mu\right)
-\frac{2}{3m^2}p_\mu p_\nu\right].
\end{equation}
 Following similar algebra, the massless propagator in the physical gauge can be written as
\begin{equation}
S_{\mu\nu}(p)=\frac{i}{\slashed p+i\varepsilon}P^{3/2}_{\mu\nu}(p).
\end{equation}

\end{document}